\documentclass[aps,pre,twocolumn,groupedaddress,showpacs]{revtex4}

\usepackage{graphicx}
\usepackage{dcolumn}
\usepackage{bm}
\usepackage{amssymb}
\usepackage{amsmath}

\usepackage{color}

\begin{document}

\title{The lengths distribution of laminar phases \\ for type-I intermittency in the presence of noise\footnote{Paper published as\\ Hramov A.E., Koronovskii A.A., Kurovskaja M.K., Ovchinnikov A.A., Boccaletti S. {\it Length distribution of laminar phases for type-I intermittency in the presence of noise}. Phys. Rev. E. {\bf 76}, 2 (2007) 026206}}

\author{Alexander~E.~Hramov$^{1}$}
\author{Alexey~A.~Koronovskii$^{1}$}
\author{Maria~K.~Kurovskaya$^{1}$}
\author{Alexey~A.~Ovchinnikov$^{1}$}
\author{Stefano Boccaletti$^{2,3}$}
\affiliation{$^1$ Faculty of Nonlinear Processes, Saratov State
University, Astrakhanskaya, 83, Saratov, 410012, Russia \\
$^2$ CNR --- Istituto dei Sistemi Complessi Via Madonna del Piano,
10 50019 Sesto Fiorentino (FI), Italy \\
$^3$ The Italian Embassy in Tel Aviv, Trade Tower, 25, Hamered
St., 68125 Tel Aviv, Israel}

\date{\today}

\begin{abstract}
We consider a type of intermittent behavior that occurs as the
result of the interplay between dynamical mechanisms giving rise
to type-I intermittency and random dynamics. We analytically
deduce the laws for the distribution of the laminar phases, with
the law for the mean length of the laminar phases versus the
critical parameter deduced earlier [PRE 62 (2000) 6304] being the
corollary fact of the developed theory. We find a very good
agreement between the theoretical predictions and the data
obtained by means of both the experimental study and numerical
calculations. We discuss also how this mechanism is expected to
take place in other relevant physical circumstances.
\end{abstract}


\maketitle

\section*{Introduction}
\label{sct:Introduction}


Intermittency is known to be an ubiquitous phenomenon in nonlinear
science. Its arousal and main statistical properties have been
studied and characterized already since long time ago, and
different types of intermittency have been classified as types
I--III~\cite{Berge:1988_OrderInChaos,Dubois:1983_IntermittencyIII},
on--off
intermittency~\cite{Platt:1993_intermittency,Heagy:1994_intermittency,%
Boccaletti:2000_IntermitLagSynchro,Hramov:2005_IGS_EuroPhysicsLetters},
eyelet
intermittency~\cite{Pikovsky:1997_EyeletIntermitt,Lee:1998:PhaseJumps,%
Boccaletti:2002_LaserPSTransition_PRL} and ring
intermittency~\cite{Hramov:RingIntermittency_PRL_2006}.

From the other side, increasing interest has been put recently in
the study of the constructive role of noise and fluctuations in
nonlinear dynamical systems. In particular, it was discovered that
random fluctuations can actually induce some degree of order
in a large variety of nonlinear systems~\cite{Pikovsky:1997_CoherenceResonance,%
Mangioni:1997_Noise,Zaikin:2000_DblStochRes}, and such  phenomena
were widely observed in relevant physical, chemical and biological
circumstances ~\cite{Neiman:2002_SynchroAndNoise,%
Zhou:2002_NoiseEnhancedPS,Zhou:PRE2003,Boccaletti:2002_LaserPSTransition_PRL}.

There are no doubts that different types of intermittent behavior
may take place in the presence of noise and fluctuations in a wide
spectrum of systems, including cases of practical interest for
applications in radio engineering, medical, physiological and
other applied sciences. It is plausible that such an interaction
would originate new types of dynamics. Therefore, the intermittent
behavior in the presence of noise has been studied by means of
Fokker-Plank equation~\cite{Hirsch:1982_Intermittency} and
adopting renormalization group
analysis~\cite{Hirsch:1982_IntermittencyPLA}, but the
characteristic relations were obtained only in the subcritical
region, where the intermittent behavior is observed both in the
presence of noise and without noise.
Recently~\cite{Kye:2000_TypeIAndNoise}, the theoretical
consideration of the intermittent behavior in the presence of
noise has been considered in the supercritical region (where
intermittency is absent in the absence of noise), with the
analytical form of the dependence of the mean length of the
laminar phases versus the critical parameter being deduced under
the assumption of the fixed reinjection probability taken in the
form of a $\delta$-function. Moreover, the found analytical law
has been verified by means of the experimental observation of the
characteristic relations of intermittency in the presence of
noise~\cite{Cho:2002_TypeINoseExpeiment,Kye:2003_TypeIIAndIIINoiseExperiment}.
At the same time, the other important statistical characteristic
of the intermittent behavior, namely, the distribution of the
laminar phase lengths, has not been obtained hitherto for the
supercritical parameter region.

In this paper we report for the first time the form of the
distribution of the lengths of the laminar phases deduced
analytically for the type-I intermittency in the presence of noise
for the region of the supercritical parameter values. The already
known dependence of the mean length of the laminar phases on the
criticality
parameter~\cite{Kye:2000_TypeIAndNoise,Cho:2002_TypeINoseExpeiment}
follows as a corollary of the carried out research. Moreover, we
prove that this dependence obtained
in~\cite{Kye:2000_TypeIAndNoise} under the assumption of the fixed
reinjection probability taken in the form of delta-function does
not depend practically on the relaminarization properties, and,
correspondingly, the obtained expression of the mean length of the
laminar phases on the criticality parameter remains correct for
different forms of the reinjection probability. The obtained
analytical distribution of the laminar phase length is verified by
means of both numerical calculations of the model system dynamics
and experimental observations.

The structure of the paper is the following. In
Sec.~\ref{sct:NoisedTypeIIntermittency} we describe the theory of
the type-I intermittency with noise and give the theoretical
predictions for the distributions of the laminar phase length. In
Sec.~\ref{sct:SampleSystems} we describe the dynamical systems
used to illustrate our conclusions. We show numerically that our
theoretical predictions are observed in the different nonlinear
systems, including coupled chaotic oscillators near the boundary
of the phase synchronization in the case of small detuning of the
natural frequencies. Finally, in Sec.~\ref{sct:Experiment} we give
the description of the experimental setup for the measurement of
the characteristics of the type-I intermittency in the presence of
noise. The final conclusions are given in
Sec.~\ref{sct:Conclusion}.

\section{The theory of the type-I intermittency in the presence of noise}
\label{sct:NoisedTypeIIntermittency}

The standard model that is used to study the type-I
intermittency~\cite{Berge:1988_OrderInChaos} is the one-parameter
quadratic map
\begin{equation}
x_{n+1}=f(x_n)=x_n+x_n^2+\epsilon \label{eq:QuadraticMap}
\end{equation}
where $\epsilon$ is a control parameter. The value of
$\epsilon_c=0$ corresponds to the saddle-node (tangential)
bifurcation when the stable and unstable points touch each other
in $x=0$ and disappear.

\begin{figure}[b]
\centerline{\includegraphics*[scale=0.45]{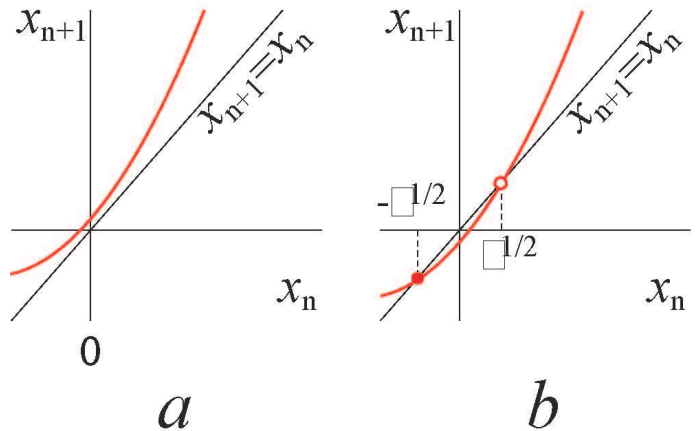}}
\caption{(Color online) The iteration diagram for
map~(\ref{eq:QuadraticMap}) (\textit{a}) $\epsilon>0$ and
(\textit{b}) $\epsilon<0$. The stable and unstable fixed points
of~(\ref{eq:QuadraticMap}) are shown by \textcolor{red}{$\bullet$}
and \textcolor{red}{$\circ$},
respectively}\label{fgr:IterationDiagram}
\end{figure}

Below the critical parameter value (i.e., for
$\epsilon<\epsilon_c$), the stable fixed point is observed, while
above $\epsilon_c$ a narrow corridor between the function $f(x)$
and the bisector $x_{n+1}=x_n$ exists, such that the point
representing the state of the map~(\ref{eq:QuadraticMap}) moves
along it (Fig.~\ref{fgr:IterationDiagram}). This movement
corresponds to the laminar phase, its mean length $T$ being
inversely proportional to the square root of
$(\epsilon-\epsilon_c)$, i.e.
\begin{equation}\label{eq:Type-IIntermittencyPowerLaw}
T\sim(\epsilon-\epsilon_c)^{-1/2}.
\end{equation}

To develop the theory of type-I intermittency in the presence of
noise, we consider the same quadratic map~(\ref{eq:QuadraticMap})
with the addition of a stochastic term $\xi_n$
\begin{equation}
x_{n+1}=x_n+x_n^2+\epsilon+\xi_n,
\label{eq:StochasticQuadraticMap}
\end{equation}
where $\xi_n$ is supposed to be a delta-correlated white noise
[${\langle\xi_n\rangle=0}$,
${\langle\xi_n\xi_m\rangle=D\delta(n-m)}$].

The influence of the stochastic term $\xi_n$ on the behavior of
the system is governed by the value of parameter $D$. For positive
values of the control parameter $\epsilon$ ($\epsilon>0$), the
point corresponding to the behavior of
system~(\ref{eq:StochasticQuadraticMap}) moves in the iteration
diagram along the narrow corridor, its motion being perturbed by
the stochastic force. As far as the intensity of the noise is not
large, the characteristics of classical type-I intermittency are
observed.

A different scenario occurs for control parameters $\epsilon$
assuming negative values ($\epsilon=-\varepsilon$, where
$\varepsilon=|\epsilon|>0$). In this case, the point corresponding
to the behavior of system~(\ref{eq:StochasticQuadraticMap}) is
localized for a long time in the region
${x<x_c=\varepsilon^{1/2}}$ and its dynamics is also perturbed by
the stochastic force. As soon as the system state point arrives at
the boundary $x_c=\varepsilon^{1/2}$ due to the influence of
noise, a turbulent phase arises, though such kind of events is
very rare.

In this case, the behavior of the
map~(\ref{eq:StochasticQuadraticMap}) differs radically from the
dynamics of the system~(\ref{eq:QuadraticMap}), since the
turbulent phases are not observed for $\epsilon<0$ if there is no
noise. Therefore, such a region of negative values of the
$\epsilon$-parameter is the main subject of interest for the
type-I intermittency in the presence of noise.

Having supposed that: (i) the value of $\epsilon$ is negative and
rather small and (ii) the value of $x$ changes per one iteration
insufficiently, we can consider ${(x_{n+1}-x_{n})}$ as the time
derivative $\dot x$ and undergo from the system with discrete
time~(\ref{eq:StochasticQuadraticMap}) to the flow system, in the
same way as in the case of the classical theory of the type-I
intermittency.

Since the stochastic term is present
in~(\ref{eq:StochasticQuadraticMap}) we have to examine the
stochastic differential equation
\begin{equation}
dX=(X^2-\varepsilon)\,dt+dW \label{eq:SDE}
\end{equation}
(where $X(t)$ is a stochastic process, $W(t)$ a one-dimensional
Winner process, {$\varepsilon=|\epsilon|$}) instead of the
ordinary differential equation ${dx/dt=x^2+\epsilon}$ considered
in the classical theory of type~I intermittency.

The stochastic differential equation~(\ref{eq:SDE}) is equivalent
to the Fokker-Plank equation
\begin{equation}
\frac{\partial \rho_X(x,t)}{\partial t}=-\frac{\partial}{\partial
x}((x^2-\varepsilon)\rho_X(x,t))+\frac{D}{2}\frac{\partial^2\rho_X(x,t)}{\partial
x^2} \label{eq:FPEMinus}
\end{equation}
for the probability density $\rho_X(x,t)$ of the stochastic
process $X(t)$. Contrarily to what done in Ref.
~\cite{Kye:2000_TypeIAndNoise} (where the backward Fokker-Plank
equation was used),  we here consider the forward one that allows
us to obtain the explicit form of the distribution of the laminar
phase lengths. The chosen initial condition is
${\rho_X(x,0)=\delta(x)}$, where $\delta(\cdot)$ is a
delta-function. Such choice of the initial form of the probability
density $\rho_X(x,0)$ corresponds to the beginning of the laminar
phase, when the point representing the state of the
system~(\ref{eq:StochasticQuadraticMap}) is in the place with
coordinate $x=0$ at time $t=0$. In other words, we suppose that
the reinjection probability is a $\delta$-function
\begin{equation}
P_{in}(x)=\delta(x)
\end{equation}
and after the relaminarization process the system is always
returned to the state $x=0$. Although the reinjection probability
$P_{in}(x)$ is well-known to be important factor and should be
taken into account when the statistical properties of the
intermittent behavior are
studied~\cite{Kin:1994_NewIntermittencyCharacteristics,%
Kim:1998_IntermittencyCharacteristicsPRL}, in the considered
problem the form of the reinjection probability practically does
not influence on the distribution of the laminar phase lengths
(and the dependence of the mean laminar phase length on the
criticality parameter, respectively), as it will be shown below.

To reduce the number of the control parameters the normalization
$z=x/\sqrt{\varepsilon}$, $\tau=t{\sqrt{\varepsilon}}$ may be
used, after which Eq.~(\ref{eq:FPEMinus}) may be rewritten in the
form
\begin{equation}
\frac{\partial \rho_Z(z,\tau)}{\partial
\tau}=-\frac{\partial}{\partial
z}((z^2-1)\rho_Z(z,\tau))+\frac{D^*}{2}\frac{\partial^2\rho_Z(z,\tau)}{\partial
z^2}, \label{eq:NormFPEMinus}
\end{equation}
where $D^*=D\varepsilon^{-3/2}$,
$\rho_Z(z,\tau)=\rho_X(z{\varepsilon}^{1/2},\tau{\varepsilon}^{-1/2})$.

Since the coordinate of the system state is localized for a long
time in the region $z<z_c=1$, we suppose that the probability
density may be written in the form ${\rho_Z(z,\tau)=A(\tau)g(z)}$,
${\forall z\leq 1}$, where $A(\tau)>0$ decreases very slowly as
time increases, i.e. $dA/d\tau\approx 0$. The function $g(z)$
should also satisfy the conditions
\begin{equation}\label{eq:ConditionsForP}
g(z)>0 ~~ \forall z<1 \quad\mathrm{and}\quad
\displaystyle\int\limits_{-\infty}^{1} g(z)\,dz<\infty.
\end{equation}

Under the mentioned assumption, we  consider the ordinary
differential equation
\begin{equation}
D^*g''(z)-2((z^2- 1)g(z))'=0\label{eq:ODUSecondOrder}
\end{equation}
instead of~(\ref{eq:NormFPEMinus}) for the region $z<1$.

This equation is equivalent to
\begin{equation}
D^*g'(z)-2(z^2- 1)g(z)+C_1=0, \label{eq:ODUFirstOrder}
\end{equation}
where $C_1$ is constant. To solve this equation we use the
integrating factor
\begin{equation}
M(z)=\exp\left(-\frac{2}{D^*}\left(\frac{z^3}{3}- z\right)\right).
\end{equation}
The solution of~(\ref{eq:ODUFirstOrder}) may be found in the form
\begin{equation}\label{eq:SolutionOfODE}
g(z)=\frac{\displaystyle
C_1\int_0^z\exp\left(\frac{2}{D^*}\left(s-\frac{s^3}{3}\right)\right)\,ds+C_2}{\displaystyle
D^*\exp\left(\frac{2}{D^*}\left(z-\frac{z^3}{3}\right)\right)}.
\end{equation}
Since the obtained $g(z)$ satisfies the
conditions~(\ref{eq:ConditionsForP}) only if $C_1\equiv 0$, the
probability density $\rho_Z(z,\tau)$ in the region $z<1$ is
\begin{equation}
\rho_Z(z,\tau)\simeq
A(\tau)\exp\left(-\frac{2}{D^*}\left(z-\frac{z^3}{3}\right)\right).
\end{equation}

The decrease of $A(\tau)$ should be determined by the probability
distribution taken in the boundary point $z=1$, i.e.,
$dA(\tau)/d\tau\sim -\rho_Z(1,\tau)$. This assumption may be
rewritten as
\begin{equation}
\frac{dA(\tau)}{d\tau}=-kA(\tau)\exp\left(-\frac{4}{3D^*}\right),
\end{equation}
where $k$ is a proportionality coefficient. Evidently, the
decrease of $A(\tau)$ is described by the exponential law
\begin{equation}
A(\tau)=A(0)\exp(-k\eta\tau),\quad \eta=\exp{(-4/(3D^*))}.
\end{equation}

Having returned to the initial variables $x$ and $t$ we obtain the
following expression for the probability density $\rho_X(x,t)$
\begin{equation}\label{eq:ProbabDistribMinus}
\rho_X(x,t)\simeq A(t)\exp\left(-\frac{2}{D}\left(\varepsilon
x-\frac{x^3}{3}\right)\right),
\end{equation}
where
\begin{equation}\label{eq:AvsT}
A(t)=A(0)\exp\left(-\frac{t}{T}\right),
\end{equation}
and
\begin{equation}\label{eq:TheoreticalLawForMeanLenght}
T=\frac{1}{k\sqrt{\varepsilon}}\exp\left(\frac{4\varepsilon^{3/2}}{3D}\right),
\end{equation}
with $A(t)$ being considered as a normalizing factor, i.e.,
\begin{equation}\label{eq:NormalizationConditionForA}
A(0)\int\limits_{-\infty}^{\sqrt{\varepsilon}}
\exp\left(-\frac{2}{D}\left(\varepsilon
x-\frac{x^3}{3}\right)\right)\,dx=1.
\end{equation}

To confirm the assumptions made above and the obtained equations,
we have compared the evolution of the probability density
$\rho_X(x,t)$ given by~(\ref{eq:ProbabDistribMinus}) with the
result of the direct numerical calculation of the Fokker-Plank
equation~(\ref{eq:FPEMinus}) with the values of control parameters
$\varepsilon=10^{-2}$, $D=2.5\times10^{-4}$.

\begin{figure}[tb]
\centerline{\includegraphics*[scale=0.4]{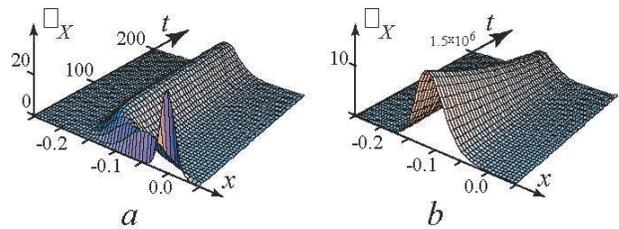}}
\caption{(Color online) The evolution of the probability density
$\rho_X(x,t)$ obtained by means of the direct numerical
integration of Fokker-Plank equation~(\ref{eq:FPEMinus}),
${\varepsilon=10^{-2}}$, ${D=2.5\times10^{-4}}$. (\textit{a}) The
initial fragment of the density evolution involving the transient
(${0\leq t<t_{tr}}$, $t_{tr}\approx 30$). (\textit{b}) The
long-time evolution of $\rho_X(x,t)$, with the transient being
omitted, $t\geq 50$} \label{fgr:ProbDensityMinusEvolution}
\end{figure}

The evolution of the probability density $\rho_X(x,t)$ obtined by
the numerical calculation of~(\ref{eq:FPEMinus}) is shown in
Fig.~\ref{fgr:ProbDensityMinusEvolution}. One can see that after
the very short transient ${0\leq t\le t_{tr}}$ the probability
density $\rho_X(x,t)$ arrives the state being close to stationary
(Fig.~\ref{fgr:ProbDensityMinusEvolution}\,\textit{a}). After that
the value of $\rho_X(x,t)$ decreases very slowly (according to the
exponential law) with time increasing, with the form of the
dependence of the probability density on $x$-coordinate being
invariable (Fig.~\ref{fgr:ProbDensityMinusEvolution}\,\textit{b}).

\begin{figure}[tb]
\centerline{\includegraphics*[scale=0.25]{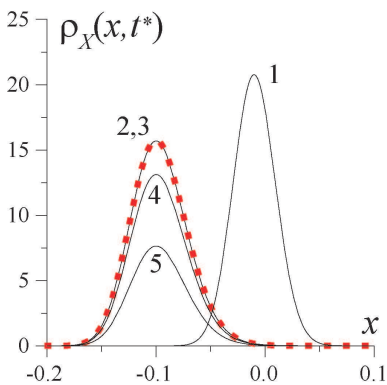}}
\caption{(Color online) The profiles of the probability density
$\rho_X(x,t^*)$ taken in the different moments of time $t^*$
obtained from the direct numerical calculation of Fokker-Plank
equation~(\ref{eq:FPEMinus}). }
\label{fgr:ProbDensityMinusProfiles}
\end{figure}

Fig.~\ref{fgr:ProbDensityMinusProfiles} also shows the profiles of
the probability density $\rho_X(x,t^*)$ taken in the different
moments of time. It is evident, that after a very short transient
(curve~1, $t_1^*=10$), the density $\rho_X(x,t)$ practically does
not change when time increases.

Two different profiles $\rho_X(x,t^*)$ corresponding to the time
moments $t^*_2=3\times10^1$ and $t^*_3=2\times10^4$ (curves~2 and
3, respectively) are very close to each other despite of the very
large time interval $\Delta t={t^*_3-t^*_2}$ between them.
Moreover, they are in  very good agreement with the approximated
solution $A(0)g(x)$ described by Eq.~(\ref{eq:ProbabDistribMinus})
and shown in Fig.~\ref{fgr:ProbDensityMinusProfiles} by means of
 squares. As time goes on, the amplitude of the probability density decreases according
to the exponential law, but very slowly (see
Fig.~\ref{fgr:ProbDensityMinusProfiles}, curves~4 and 5,
$t^*_4=2.5\times10^5$ and $t^*_5=10^6$, respectively), although
the probability density form remains the same for all  times.

Therefore, taking into account the results of the direct numerical
calculations of Fokker-Plank equation~(\ref{eq:FPEMinus}) and the
comparison with the obtained approximated
solution~(\ref{eq:ProbabDistribMinus}), we come to  the conclusion
that our assumptions  are correct and can be used for the further
analysis.

The evolution of the probability density $\rho_X(x,t)$ may be
considered separately on two time intervals ${0\leq t < t_{tr}}$
and ${t_{tr}\leq t < +\infty}$, respectively. The first time
interval corresponds to the transient when the probability density
$\rho_X(x,t)$ evolves to the form~(\ref{eq:ProbabDistribMinus})
being close to stationary. Only when ${0\leq t < t_{tr}}$ the form
of the reinjection probability $P_{in}(x)$ may influence on the
evolution of the probability density $\rho_X(x,t)$. For  $t\geq
t_{tr}$ (when the transient is elapsed), the evolution of the
probability density is defined completely by
Eq.~(\ref{eq:ProbabDistribMinus}) and it does not depend entirely
on the reinjection probability $P_{in}(x)$. Since the transient is
very short in comparison with the exponential decrease of the
probability density $\rho_X(x,t)$ we can neglect them and use only
the second time interval ${t_{tr}\leq t < +\infty}$ to obtain the
statistical characteristics of the type-I intermittent behavior in
the presence of noise. It is clear, that in this case the obtained
results do not depend on the relaminarization process and the
reinjection probability $P_{in}(x)$.

The distribution $p(t)$ of the laminar phase lengths $t$ may be
defined from the relationship between $\rho_X(x,t)$ and $p(t)$
\begin{equation}
p(t)=-\int\limits_{-\infty}^{\sqrt{\varepsilon}}\frac{\partial\rho_X(x,t)}{\partial
t}\,dx.
\end{equation}
Using relations~(\ref{eq:ProbabDistribMinus}), (\ref{eq:AvsT}) and
(\ref{eq:NormalizationConditionForA}) one can obtain, that the
laminar phase distribution is governed by the exponential law
\begin{equation}\label{eq:LamPahseLengthDistribution}
p(t)=T^{-1}\exp\left(-{t/T}\right),
\end{equation}
where $T$ defined by Eq.~(\ref{eq:TheoreticalLawForMeanLenght}) is
the mean length of the laminar phases. The obtained
expression~(\ref{eq:TheoreticalLawForMeanLenght}) for the mean
length $T$ of the laminar phases is consistent with the formal
solution $T\sim\varepsilon^{-1/2}f(\sigma^2\varepsilon^{-3/2})$
derived in the previous
considerations~\cite{Hirsch:1982_Intermittency,Hirsch:1982_IntermittencyPLA,Crutchfield:1982_Fluctuations}
and coincides with the formula of $T$ given
in~\cite{Kye:2000_TypeIAndNoise}. If the criticality parameter
$\varepsilon$ is large enough, the approximate equation ${\ln
T\sim D^{-1}\varepsilon^{3/2}}$ may be used (see
\cite{Kye:2000_TypeIAndNoise} for detail). Based on the
consideration carried out above we state that
Eqs.~(\ref{eq:TheoreticalLawForMeanLenght}) and
(\ref{eq:LamPahseLengthDistribution}) do not depend practically on
the relaminarization process properties and may be used for the
arbitrary reinjection probability $P_{in}(x)$.

\section{Sample system dynamics}
\label{sct:SampleSystems}

To verify the obtained theoretical predictions, we consider
numerically two different dynamical systems showing type-I
intermittency, with a stochastic force being added. As such test
systems we have selected (i) the quadratic map and (ii) driven Van
der Pol oscillator.

\subsection{Quadratic map with stochastic force}
\label{sbsct:QMap}

We start considering the interplay between type-I intermittency
and noise using the quadratic map
\begin{equation}
x_{n+1}=x_n^2+\lambda+\epsilon+D\xi_n,\mod 1,
\label{eq:TestQuadraticMap}
\end{equation}
where the  ``$\mathrm{mod~}1$'' operation is used to provide the
return of the system in the vicinity of the point $x=0$;
$\lambda=0.25$, and the probability density of the stochastic
variable $\xi$ is distributed uniformly on the interval
$\xi\in[-1,1]$. The map~(\ref{eq:TestQuadraticMap}) may be brought
to~(\ref{eq:StochasticQuadraticMap}) with the help of a linear
variable transformation. If the intensity of noise $D$ is equal to
zero the saddle-node bifurcation is observed for $\epsilon=0$. The
type-I intermittent behavior is observed for $\epsilon>0$, whereas
the stable fixed point takes place for $\epsilon<0$. Having added
the stochastic force $D=10^{-7}$ in~(\ref{eq:TestQuadraticMap}) we
can expect that the intermittent behavior may be also observed in
the area of the negative values of the criticality parameter
$\epsilon$.

\begin{figure}[tb]
\centerline{\includegraphics*[scale=0.45]{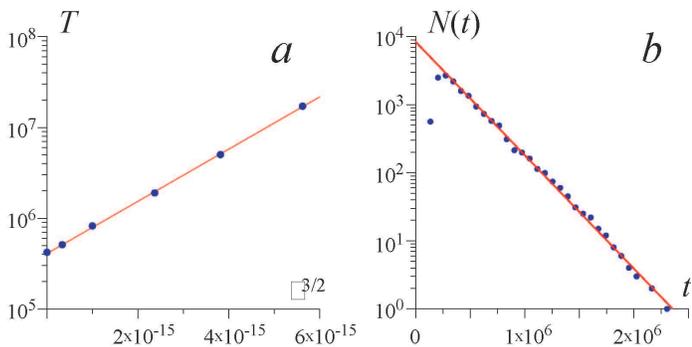}} \caption{(Color
online) (\textit{a}) The dependence of the mean length $T$ of the
laminar phases on the criticality parameter $\varepsilon=|\epsilon|$
($\epsilon<0$) for~(\ref{eq:TestQuadraticMap}). The points obtained
by the iteration of~(\ref{eq:TestQuadraticMap}) are shown by symbols
(\textcolor{blue}{$\bullet$}). The theoretical law ${\ln T\sim
\varepsilon^{3/2}}$ is shown by the solid line. (\textit{b}) The
distribution of the laminar phase lengths for
map~(\ref{eq:TestQuadraticMap}), the criticality parameter value has
been selected as $\varepsilon=10^{-11}$. The theoretical exponential
law~(\ref{eq:LamPahseLengthDistribution}) is shown by the solid
line} \label{fgr:TestQuadraticMapMeanLengthMinus}
\end{figure}

The dependence of the mean laminar phase length $T$ on the
criticality parameter below the point $\epsilon_c=0$ is shown in
Fig.~\ref{fgr:TestQuadraticMapMeanLengthMinus}. To compare it with
the obtained theoretical prediction ${\ln T\sim
D^{-1}\varepsilon^{3/2}}$ the abscissa in
Fig.~\ref{fgr:TestQuadraticMapMeanLengthMinus},\,\textit{a} has
been selected in the $\varepsilon^{3/2}$-scale
($\varepsilon=|\epsilon|$, $\epsilon<0$), whereas the ordinate
axis is shown in the logarithmic scale. One can see the excellent
agreement between theoretical
law~(\ref{eq:TheoreticalLawForMeanLenght}) and data of numerical
calculation.

The distribution of the laminar phase lengths  is also in  very
good accordance with the exponential
law~(\ref{eq:LamPahseLengthDistribution}) predicted by the theory
of the type-I intermittency with noise (see
Fig.~\ref{fgr:TestQuadraticMapMeanLengthMinus},\,\textit{b}). Note
the presence of the small region of the short laminar phase
lengths in
Fig.~\ref{fgr:TestQuadraticMapMeanLengthMinus},\,\textit{b} where
the deviation from the prescribed exponential
law~(\ref{eq:LamPahseLengthDistribution}) is observed. This region
corresponds to the transient ${0\leq t < t_{tr}}$ when the
probability density $\rho_X(x,t)$ evolves to the
form~(\ref{eq:ProbabDistribMinus}) being close to stationary as it
was discussed in Sec.~\ref{sct:NoisedTypeIIntermittency}. The
existence of this transient time interval does not influence
practically on the
characteristics~(\ref{eq:TheoreticalLawForMeanLenght}) and
(\ref{eq:LamPahseLengthDistribution}) of the intermittent behavior
in the presence of noise in the full agreement with the
conclusions made above.

So, the intermittent behavior observed in the quadratic map with
the stochastic force agrees well with the theoretical predictions
obtained in Sec.~\ref{sct:NoisedTypeIIntermittency}. Since the
theory of type-I intermittency has been developed on the basis of
the model~(\ref{eq:StochasticQuadraticMap}) being very close to
map~(\ref{eq:TestQuadraticMap}), it is mandatory to examine
another system to ensure that our theoretical conclusions are
correct and applicable for a wide spectrum of  nonlinear systems.

\subsection{Van der Pol oscillator driven by the external harmonic signal in the presence of noise}
\label{sbsct:VdPSinAndNoise}

We consider as a second model the system given by a van der Pol
oscillator
\begin{equation}
{\ddot{x}-(\lambda-x^2)\dot{x}+x=A\sin(\omega_et)+D\xi(t)}
\label{eq:DrivenVdPOscillatorAndNoise}
\end{equation}
driven by an external harmonic signal with the amplitude $A$ and
frequency $\omega_e$, with an added stochastic term $D\xi(t)$. The
values of the control parameters are selected to be $\lambda=0.1$,
$\omega_e=0.98$. For these control parameters and for $D=0$, the
dynamics of the driven van der Pol oscillator becomes synchronized
when $A=A_c=0.0238$. The probability density of the random
variable $\xi(t)$ is
\begin{equation}
p(\xi)=\frac{1}{\sqrt{2\pi}\sigma}\exp\left(-\frac{\xi^2}{2\sigma^2}\right),
\label{eq:NormalDistrib}
\end{equation}
where $\sigma^2=1$. To integrate
Eq.~(\ref{eq:DrivenVdPOscillatorAndNoise}) the one-step Euler
method has been used with time step $h=5\times10^{-4}$, the value
of the noise intensity has been fixed as $D=1$.

It is well-known that under certain conditions (i.e., for the
periodically forced weakly nonlinear isochronous oscillator), the
complex amplitude method may be used to find the solution
describing the behavior of
oscillator~(\ref{eq:DrivenVdPOscillatorAndNoise}) without noise in
the form ${x(t)=\mathrm{Re}\,a(t)e^{i\omega t}}$. For the complex
amplitude $a(t)$ one obtains the averaged (truncated) equation
${\dot{a}=-i\nu a+a-|a|^2a-ik}$, where ${\nu}$ is the frequency
mismatch, and $k$ is the (re-normalized) amplitude of the external
force. For the small $\nu$ and large $k$, the stable fixed point
on the plane of the complex amplitude ${a^*=\mathrm{const}}$
corresponds to the synchronous regime, with the synchronization
destruction corresponding to the saddle-node bifurcation
associated with the global bifurcation of the limit cycle
birth~\cite{Pikovsky:2000_SynchroReview,Hramov:2007_2TypesPSDestruction}.
Therefore, below the boundary of the synchronization regime (for
 small values of the frequency mistuning), the dynamics of the
phase difference
\begin{equation}
\Delta\varphi(t)=\varphi(t)-\omega_et
\end{equation}
(where $\varphi(t)$ is the phase of the driven oscillator)
demonstrates time intervals of phase synchronized motion (laminar
phases) persistently and intermittently interrupted by phase slips
(turbulent phases) during which the value of $|\Delta\varphi(t)|$
jumps up by $2\pi$. The mean length $T$ of the laminar
(synchronous) phases depends on the criticality parameter
${\epsilon=(A_c-A)}$ according to the power
law~(\ref{eq:Type-IIntermittencyPowerLaw}) corresponding to the
type-I intermittency.

\begin{figure}[tb]
\centerline{\includegraphics*[scale=0.45]{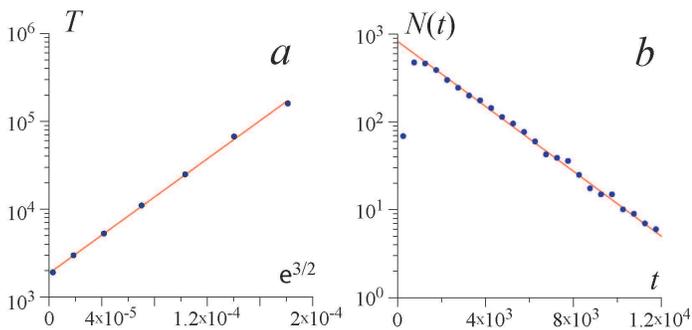}} \caption{(Color
online) (\textit{a}) The dependencies of the mean length $T$ of the
laminar phases on the criticality parameter ${\epsilon=(A_c-A)}$ for
the driven van der Pol oscillator with the stochastic
force~(\ref{eq:DrivenVdPOscillatorAndNoise}). The points obtained by
the numerical integration of~(\ref{eq:DrivenVdPOscillatorAndNoise})
are shown by symbols \textcolor{blue}{$\bullet$}. The theoretical
law ${\ln T\sim \varepsilon^{3/2}}$ is shown by the solid line.
(\textit{b}) The distribution of the laminar phase lengths for the
driven van der Pol oscillator, with the amplitude of the external
signal $A=0.0245$ being taken above the critical point $A_c=0.0238$
($\epsilon=-7\times10^{-4}$). The ordinate axis is presented in the
logarithmic scale. The theoretical exponential
law~(\ref{eq:LamPahseLengthDistribution}) is shown by the solid
line} \label{fgr:VdPMeanLength}
\end{figure}

If the stochastic term $D\xi(t)$ is added ($D\neq0$) the
manifestation of the regularities of type-I intermittency with
noise described above is revealed in the parameter range ${A>A_c}$
(see Fig.~\ref{fgr:VdPMeanLength},\,\textit{a}). For the negative
values of the criticality parameter $\epsilon$ the law ${\ln T\sim
D^{-1}\varepsilon^{3/2}}$ is expected to be observed. To make this
law evident, the abscissa in
Fig.~\ref{fgr:VdPMeanLength},\,\textit{a} has been selected in the
$\varepsilon^{3/2}$-scale ($\varepsilon=|\epsilon|$) and the
ordinate axis $T$ is shown in the logarithmic scale. One can see
again the excellent agreement between the numerically calculated
data and theoretical
prediction~(\ref{eq:TheoreticalLawForMeanLenght}).

The distribution of the lengths of the laminar phases $N(t)$
obtained for $A>A_c$ also confirms the theoretical results given
above. Indeed, the distribution $N(t)$ shown in
Fig.~\ref{fgr:VdPMeanLength},\,\textit{b} is in the very good
accordance with the theoretically predicted exponential
law~(\ref{eq:LamPahseLengthDistribution}), with the small region
of the short laminar phase lengths deviating from the exponential
law being revealed as well as for the discrete
map~(\ref{eq:TestQuadraticMap}) that corresponds to the short
transient taking place after the relaminarization. Again, as for
the discrete map considered in Sec.~\ref{sbsct:QMap}, the
existence of this transient time interval does not distort the
characteristics of the intermittent behavior observed in the
presence of noise.

\section{Experimental observation of the characteristics of type-I intermittency in the presence of noise}
\label{sct:Experiment}

In parallel with the numerical analysis of the type-I intermittent
behavior with noise we have also studied experimentally the
dynamics of the periodic oscillator driven by the the external
harmonic signal in the presence of noise to confirm the
theoretical and numerical results given in
Sec.~\ref{sct:NoisedTypeIIntermittency} and
\ref{sct:SampleSystems}. In the experiment we have used the simple
electronic oscillator where all parameters (including noise
amplitude) may be controlled precisely.

\begin{figure}[tb]
\centerline{\includegraphics*[scale=0.38]{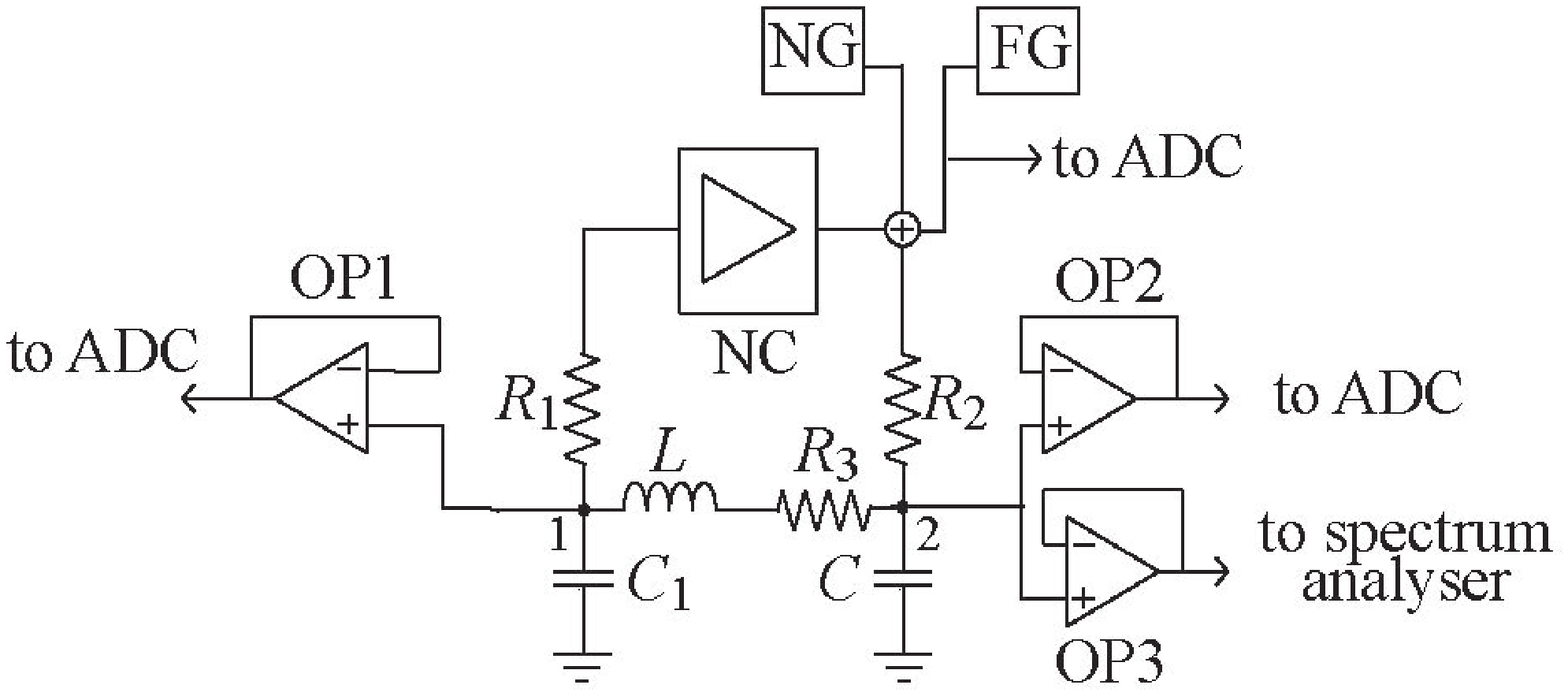}} \caption{The
schematic diagram of the experimental setup. The control parameters
have been selected as the following: $R_1=10$~Ohm, $R_2=630$~Ohm,
$R_3=56$~Ohm, $L=3.3$~mH, $C=150$~nF, $C_1=330$~nF. The operational
amplifiers OP1 and OP2 are both of the TL082 type and the
operational amplifier OP3 is of the TDA2030 type}
\label{fgr:SetupScheme}
\centerline{\includegraphics*[scale=0.4]{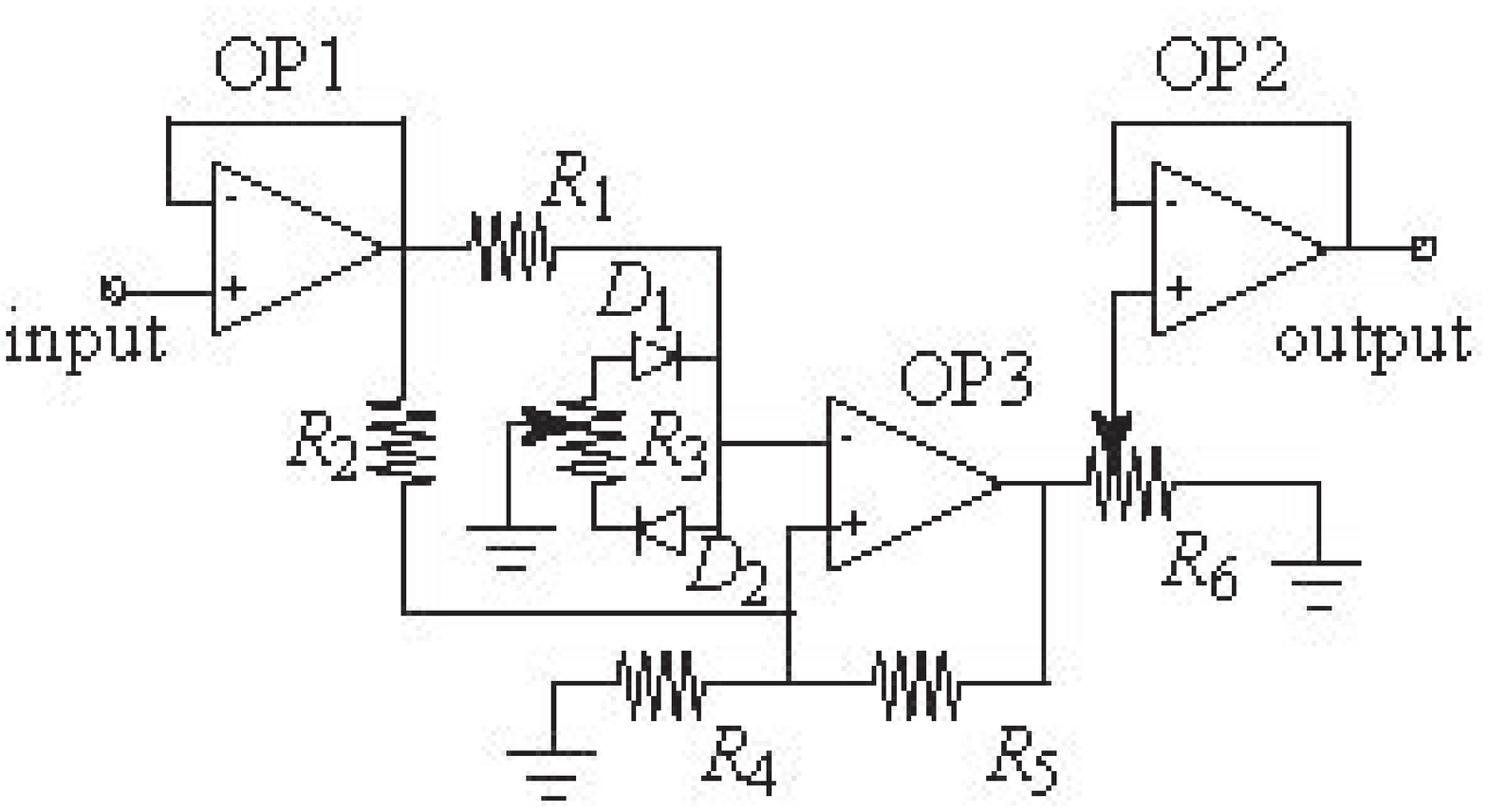}} \caption{The
schematic diagram of the nonlinear converter. The control parameters
have been selected as the following: $R_1=2.7$~kOhm, $R_2=7.5$~kOhm,
$R_3=100$~Ohm, $R_4=7.5$~kOhm, $R_5=12$~kOhm, $R_6=4.7$~kOhm. The
diodes $D_1$ and $D_2$ are of the 1N4148 type. The operational
amplifiers OP1 and OP2 are both of the TL082 type and the
operational amplifier OP3 is of the LF356N type}
\label{fgr:NCScheme}
\end{figure}

The experimental setup is shown in Fig.~\ref{fgr:SetupScheme}. The
basis element of the scheme we use the generator with the linear
feedback and nonlinear converter
(NC)~\cite{Rulkov:1996_SynchroCircuits}. The diagram of the
nonlinear converter is shown in Fig.~\ref{fgr:NCScheme}. The
characteristics of nonlinear converter were controlled with
resistor $R_6$ (see Fig.~\ref{fgr:NCScheme}). Since the generator
demonstrates both chaotic and periodic oscillations, the control
parameters of it have been selected in such a way for the
generated signal to be periodic. The frequency of the autonomous
periodic oscillations was $8.805$~kHz. As a source of driving
harmonic signal the MOTECH-FG503 functional generator (FG) has
been used. The behavior of the oscillator driven by the external
harmonic signal in the presence of noise was analyzed by means of
the Agilent E4402B spectrum analyzer and L-Card L-783
analog--digital converter (ADC) PCI-card with 12-bit resolution.

\begin{figure}[tb]
\centerline{\includegraphics*[scale=0.425]{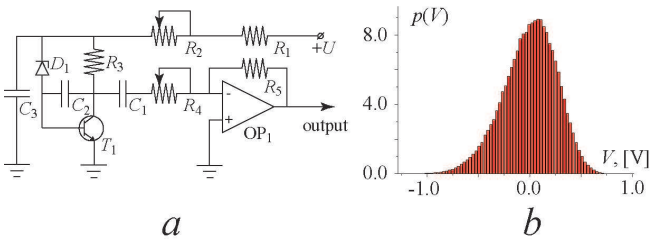}}
\caption{(Color online) (\textit{a}) The schematic diagram of the
noise generator. The control parameters have been selected as the
following: $R_1=72$~Ohm, $R_2=1$~kOhm, $R_3=1$~kOhm, $R_4=10$~kOhm,
$R_5=10$~kOhm, $C_1=1$~$\mu$F, $C_2=0.1$~$\mu$F, $C_3=470$~$\mu$F,
$U=15$~V. The diode $D_1$ is of the BZX79-C11 type. The operational
amplifier OP1 is of the TL082 type, the transistor $T_1$ is of the
BC546B type. (\textit{b}) The distribution $p(V)$ of noise}
\label{fgr:NGScheme}
\end{figure}

The noise generator (NG) shown in
Fig.~\ref{fgr:NGScheme},\,\textit{a} provides the noise signal
being close to Gaussian
one~\cite{Sze:1981_SemiconductorDevicesBook}. The distribution of
noise $p(V)$ is shown in Fig.~\ref{fgr:NGScheme},\,\textit{b}. The
intensity of noise may be controlled easily by means of the
variation of the potentiometer $R_4$.

As well as for Van der Pol oscillator driven by the external
harmonic signal in the presence of noise (see
Sec.~\ref{sbsct:VdPSinAndNoise}) below the boundary of the
synchronization regime (for
 small values of the frequency mistuning) the dynamics of the
difference
\begin{equation}
\Delta\varphi(t)=\varphi(t)-\varphi_e(t)
\end{equation}
between the phase of the driven oscillator $\varphi(t)$ and the
phase of the external harmonic signal $\varphi_e(t)$ should
demonstrate time intervals of phase synchronized motion (laminar
phases) persistently and intermittently interrupted by phase slips
(turbulent phases) during which the value of $|\Delta\varphi(t)|$
jumps up by $2\pi$. The distribution of the laminar phase length
is expected to obey to the exponential
law~(\ref{eq:LamPahseLengthDistribution}) predicted by the theory
of the type-I intermittency with noise.

\begin{figure}[tbh]
\centerline{\includegraphics*[scale=0.35]{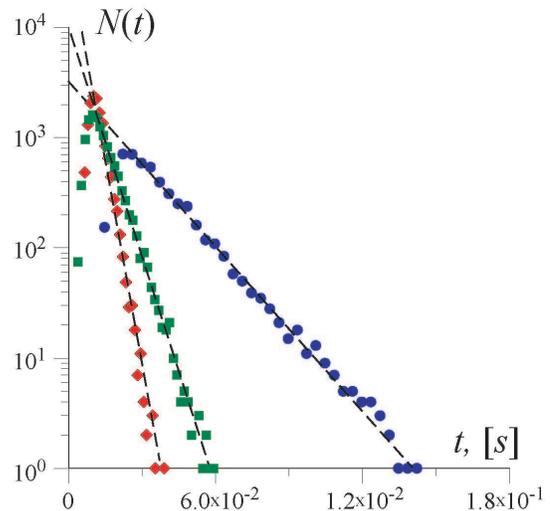}} \caption{(Color
online) The distributions of the laminar phase lengths for the
driven periodic oscillator in the presence of noise obtained
experimentally. The amplitude $V_m$ of the external signal and the
noise dispersion $\sigma$ have been selected the following:
(\textcolor{red}{$\blacklozenge$})~${V_m=170}$~mV,
${\sigma=475.17}$~mV;
(\textcolor{green}{$\blacksquare$})~${V_m=170}$~mV,
${\sigma=298.11}$~mV; (\textcolor{blue}{$\bullet$})~${V_m=150}$~mV,
${\sigma=141.72}$~mV. The ordinate axis is presented in the
logarithmic scale. The approximations corresponding to the
theoretical exponential law~(\ref{eq:LamPahseLengthDistribution})
are shown by the dashed lines } \label{fgr:ExperimentalDistribs}
\end{figure}

Since the dependence of the mean laminar phase length on the
criticality parameter has already been studied
experimentally~\cite{Cho:2002_TypeINoseExpeiment} in our
experiment we focus on the consideration of distribution of the
laminar phase lengths. These distributions $N(t)$ obtained
experimentally for the different values of the amplitude $V_m$ of
the external harmonic signal and noise intensity $D$ are shown in
Fig.~\ref{fgr:ExperimentalDistribs}. The frequency of the external
harmonic signal has been fixed as $f=8.75$~kHz, the values of the
amplitude $V$ of the external signal have been selected in such a
way for the driven oscillator to be synchronized if the intensity
of noise is equal to zero, i.e., $V_m>V_c$, where $V_c=145$~mV is
the amplitude of the external signal corresponding to the
synchronization threshold. In the presence of noise the phase
slips are revealed and the intermittent behavior is observed. One
can see that the distributions of the lengths of the laminar
phases $N(t)$ are in the very good accordance with the
theoretically predicted exponential
law~(\ref{eq:LamPahseLengthDistribution}). The small regions of
the short laminar phase lengths where the deviation from the
prescribed exponential law~(\ref{eq:LamPahseLengthDistribution})
is observed also take place as well as in the case of the
numerical simulations of the model systems considered in
Sec.~\ref{sct:SampleSystems}. Therefore, we come to conclusion
that the experimental observations confirm the obtained
theoretical results concerning the type-I intermittent behavior in
the presence of noise.

\section{Conclusions}
\label{sct:Conclusion}

In conclusion, we have reported  a  type of intermittency behavior
caused by the cooperation between the deterministic mechanisms and
random dynamics. Having examined three sample systems both
numerically and experimentally we can conclude that (i) noise
induces new features in the intermittent behavior of a system
demonstrating type-I intermittency, with  new dynamical properties
being observed above the former value of the criticality
parameter; (ii)~the results of numerical simulations and
experimental observations are in excellent agreement with the
developed theory; (iii)~the statistical characteristics of the
perturbations of type-I intermittency as well as the
relaminarization process properties and the reinjection
probability do not seem to play a major role. Though the
characterization of the intermittent process has been explicitly
derived here for model systems, we expect that the very same
mechanism can be observed in many other relevant circumstances
where the level of natural noise is sufficient, e.g. in the
physiological~\cite{Hramov:2006_Prosachivanie,Hramov:2007_UnivariateDataPRE}
or physical systems~\cite{Boccaletti:2002_LaserPSTransition_PRL}.

\section*{Acknowledgement}

We thank Prof.~T.E.~Vadivasova for the helpful comments. Work
partly supported by RFBR (projects 05--02--16273 and 07-02-00044),
by the Supporting program of leading Russian scientific schools
(project NSh--4167.2006.2), and by the ``Dynasty'' Foundation.

\bibliographystyle{apsrev}

\end{document}